\documentclass[aps,prl,superscriptaddress,reprint]{revtex4-2}
\usepackage{graphicx}
\usepackage{amsmath}
\usepackage{amssymb}
\usepackage{color}

\begin{document}

\title{A strain-engineered graphene qubit in a nanobubble}
\author{Nojoon Myoung}
\affiliation{Department of Physics Education, Chosun University, Gwangju 61452, Republic of Korea}
\author{JungYun Han}
\author{Hee Chul Park}
\email{correspondence to hcpark@ibs.re.kr}
\affiliation{Center for Theoretical Physics of Complex Systems, Institute for Basic Science, Daejeon 34051, Republic of Korea}
\affiliation{Basic Science Program, Korea University of Science and Technology (UST), Daejeon 34113, Republic of Korea}
\date{\today}

\begin{abstract}
We propose a controllable qubit in a graphene nanobubble with emergent two-level systems induced by pseudo-magnetic fields. We found that double quantum dots can be created by the strain-induced pseudo-magnetic fields of a nanobubble, and that their quantum states can be manipulated by either local gate potentials or the pseudo-magnetic fields. Graphene qubits clearly exhibit an avoided crossing behavior via electrical detuning, with energy splittings of about a few meV. We also show a remarkable tunability of our device design that allows for the fine control of the Landau--Zener transition probability through strain engineering of the nanobubble, showing half-and-half splitting at the avoided crossing point. Further, we demonstrate that the two-level systems in the nanobubble exhibit Rabi oscillations near the avoided crossing point, resulting in very fast Rabi cycles of a few ps.
\end{abstract}

\maketitle

Graphene is an attractive candidate for qubits due to its long relaxation and coherent times~\cite{Huertas-Hernando2006,TRAUZETTEL2007,Borysenko2010,Han2012}. Since a qubit consists of two quantum states occupying discrete levels, the formation of graphene quantum dots is required to construct graphene qubits. Although there have been a number of studies on graphene quantum dots (QDs) for qubit applications~\cite{TRAUZETTEL2007,Pedersen2008,Recher2009,Recher2010,Wu2012,Wang2015,Chen2015,Dong2021}, the major challenge of the QD-based graphene qubit is the material's lack of a band gap, which does not allow for Dirac fermion confinement through electrical gating. As approaches to tackle this issue, physically etched graphene nanoconstrictions~\cite{Guttinger2008,Guttinger2009,Ihn2010,Guttinger2012,Engels2013}, graphene nanoribbons~\cite{Stampfer2009,Gallagher2010,Wang2017}, and bilayer graphene~\cite{Pereira2007,daCosta2014,daCosta2015,daCosta2016,Velasco2018,Eich2018,Kurzmann2019a,Kurzmann2019b,Banszerus2020} have been investigated to achieve the bound states of Dirac fermions in graphene QD, relying on the fabrication of double quantum dots. However, compared to pristine graphene, these efforts for gap opening inevitably lead to unwanted disorders~\cite{Bischoff2012,Engels2013} that interrupt the precise control of the electronic states of individual QD~\cite{Banszerus2020}.

With the experimental confirmation of strain-induced Landau level formation in graphene~\cite{Levy2010}, there have been advancements in both theory and experimental technology, making room for unraveling the strain engineering of graphene~\cite{Pereira2009a,Guinea2010a,Guinea2010b,Vozmediano2010,Low2010,Klimov2012,Xu2012,Zhu2014}. The generation of pseudo-magnetic fields from the elastic strain in graphene has attracted enormous research interest in the tunability of the transport properties of graphene by strain. In particular, there have been theoretical expectations that Dirac fermions can be localized by non-uniform pseudo-magnetic fields~\cite{Qi2014,Bahamon2015,Myoung2020}. Such a fact, that a nanobubble can host localized states in graphene, allows for the possibility of graphene-based QD devices like qubits.

Another aspect of graphene is that its mechanical properties are conducive to the study of electromechanical phenomena in nanoscale devices. Not only is graphene a rigid material due to its high elastic modulus reaching up 1 TPa~\cite{Lee2008}, but it is also a stretchable material with a large failure strength exceeding 100 GPa~\cite{Liu2007,Cadelano2009,Zhao2009}. Since graphene exhibits a wide range of elastic strain up to 20\%~\cite{Cadelano2009,Zhao2009,Wei2009}, researchers have discussed the strain engineering of graphene for nanoelectromechanical devices with the potential ability to tune the electrical properties via strain control~\cite{Pereira2009a,Choi2010,Cocco2010,Guinea2010a,Gui2008,Mohr2009,Ni2008,Fu2011,Zhao2015}. So far, research progress with strained graphene has centered on band-gap engineering~\cite{Ni2008,Pereira2009a} and valleytronics~\cite{Myoung2020,Hsu2020,Low2010,Wu2011,Georgi2017,Zhai2010}, while the existence of localized states as a consequence of strain-induced pseudo-magnetic fields has remained relatively overlooked. Here, we consider a combination of strain-tunability and strain-induced localized states for graphene-based qubit applications.

In this letter, we scrutinize the possibility of using strain-induced QD states in graphene for a novel type of qubit via theoretical investigation. The graphene qubit consists of two coupled localized states inside a nanobubble as a double quantum dot (DQD), resulting in a two-level system. We propose that the two-level system is electrostatically tunable via local gate control, allowing for the initialization and modulation of the qubit. The formation of a qubit is confirmed by the avoided crossing behavior of the two-level system from resonant conductance calculations through the DQD. Moreover, we show that the Landau--Zener transition can be finely controlled by the strain of the nanobubble. Such strain-tunability enables us to secure an exact half-and-half splitting ratio of the superposed quantum states at the avoided crossing point.

We consider an armchair graphene nanoribbon where a nanobubble  is formed at the center, modeled as a Gaussian-shaped vertical deformation,
\begin{align}
z\left(\vec{r}\right)=h_{0}e^{-r^{2}/2\sigma^{2}},
\end{align}
where $h_{0}$ and $\sigma$ represent the height and radial size of the nanobubble. It is noteworthy that for $B=97.8$ T, the ribbon width $W=76.8$ nm is sufficiently larger than the magnetic length $l_{B}=\sqrt{\hbar/2eB}\simeq 3.2$ nm, so that the graphene nanoribbon can be regarded as a graphene sheet. In addition, the nanobubble is characterized by defining an effective radius $R=2\sqrt{2}\sigma$, and an appropriate $\sigma=7.2$ mn is set to make $2R=40.4$ nm smaller than $W$.

It has been previously reported that a circular nanobubble produces a characteristic non-uniform pseudo-magnetic fields with $120^{\circ}$ rotational symmetry\cite{Myoung2020}. From the tight-binding point of view, lattice deformation due to such a nanobubble is reflected in the modified hopping terms, $t_{ij}=t_{0}\mbox{exp}\left(-dt_{ij}\right)$ with
\begin{align}
dt_{ij}=\beta\left(\frac{d_{ij}}{a_{0}}-1\right),
\end{align}
where $d_{ij}$ is the distance between the $i$ and $j$ sites of graphene due to the nanobubble, $t_{0}$ is the hopping energy of unstrained graphene, $\beta\sim 3.37$, and $a_{0}$ is the lattice constant of graphene. For simplicity, we consider nearest-neighbor hopping only.

\begin{figure}
\centering
\includegraphics[width = \linewidth]{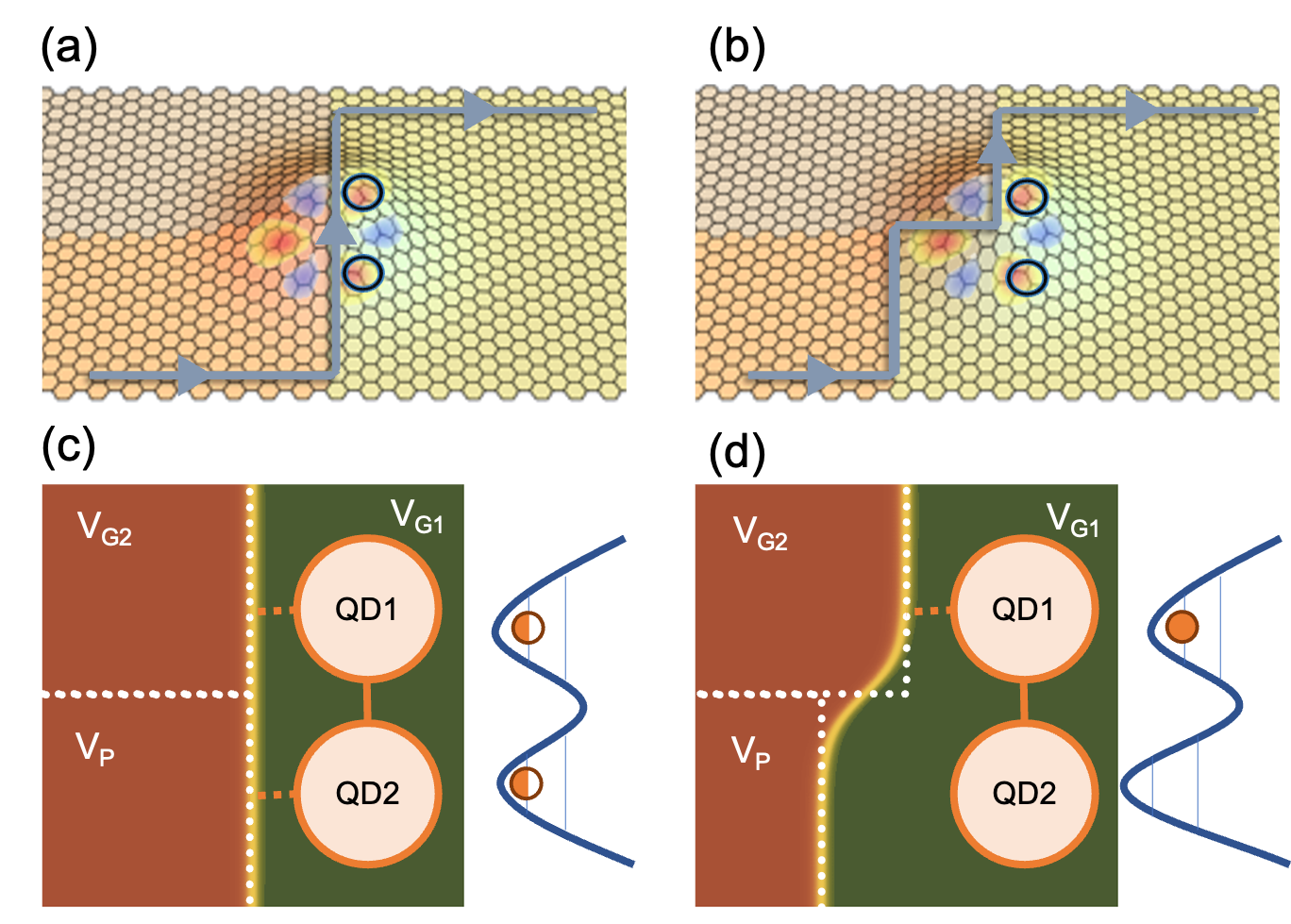}
\caption{(a) and (b) System schematics of strain-induced DQDs in graphene with symmetric and asymmetric p-n junctions, respectively. The p-n junction structures consist of three local gate electrodes $V_{G1}$, $V_{G2}$, and $V_{P}$. A quantum Hall interface channel is created along the p-n junction interface, and the electronic states of the DQDs are probed via coupling between the interface channel and each QD. (c) For the symmetric p-n junction ($V_{G2}=V_{P}$), the singlet state is composed of two identical QDs. (d) For the asymmetric p-n junction ($V_{G2}\neq V_{P}$), the two QDs are no longer identical with a different coupling strength between the interface channels and QDs, such that either QD1 or QD2 is occupied by an electron according to energy.}
\label{fig:model}
\end{figure}

Since the pseudo-magnetic fields is created at the nanobubble location, probing it typically requires one to exploit local probes like a scanning tunneling microscope, but here we use one-dimensional quantum Hall channels as a local probe. As shown in Fig. \ref{fig:model}, a p-n junction is introduced in the vicinity of the nanobubble for the propagation of Dirac fermions throughout the nanobubble. For a more realistic situation, rather than abrupt potential steps, we consider a gradually varying potential profile for the p-n junction: $U=U_{0}\tanh{\left(x/\xi\right)}$, leading to $\pm U_{0}$ in the p or n region. Note that we set $\xi=30~a_{0}\simeq 7.2$ nm and $U_{0}=\sqrt{2eB\hbar v_{F}^{2}}\simeq0.36$ eV, which is the spacing between the zeroth and first Landau levels.

The existence of QD states in the nanobubble is evidenced by Fano resonances in the conductance spectra across the p-n junction. When the Dirac fermion energy equals an energy level of the strain-induced QD, Fano resonances occur as a consequence of interference between the extended states in the interface channel and the localized states in the QD. Otherwise, we can simply see characteristic conductance oscillations via the valley-isospin rotation due to the pseudo-magnetic fields~\cite{Myoung2020}. In this letter, the coherent conductance is calculated by using S-matrix formalism based on the tight-binding approach of the \textsc{kwant} code~\cite{Groth2014}. Furthermore, we perform time-dependent simulations of the Landau-Zener transition near the avoided crossing point using \textsc{tkwant} code~\cite{Kloss2021}, with parameters extracted from numerical results of the conductance. 

\begin{figure}
\centering
\includegraphics[width = \linewidth]{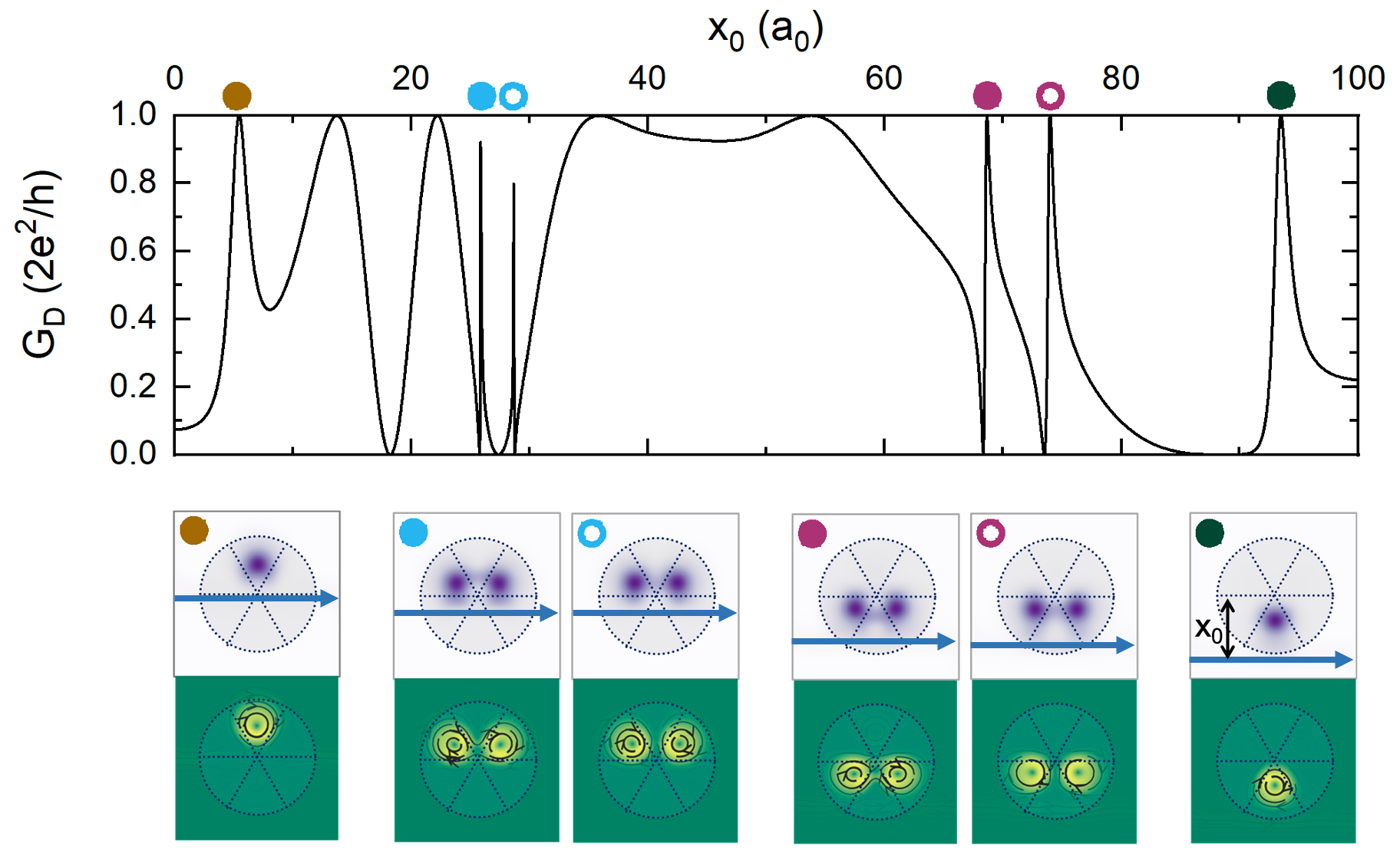}
\caption{Identification of QD formation in the nanobubble via Fano resonances. Top panel: Conductance spectrum through the nanobubble at the p-n junction interface as a function of the relative distance between the interface and the nanobubble, $x_{0}$. Fano resonances are indicated with solid and open circles, corresponding to the QD states in the nanobubble. Bottom panels: Probability and current densities of the different quantum dot states indicated by the circles with corresponding colors. The dotted lines are eye-guides for the non-uniform pseudo-magnetic field profile due to the nanobubble.}
\label{fig:QDform}
\end{figure}

In order to discuss the qubit formation, it is necessary to identify the localized states in the nanobubble. Here, to probe the localized states, we calculate the diagonal conductance $G_{D}$ across the p-n junction for $h_{0}=4.08$ nm by varying the nanobubble position $x_{0}$. As shown in Fig. \ref{fig:QDform}, the conductance exhibits a characteristic oscillation as $x_{0}$ varies as a consequence of valley-isospin rotation due to the strain-induced Berry's phase~\cite{Myoung2020}. Meanwhile, it is significant that there are a number of Fano resonances in the conductance spectra for different $x_{0}$, implying a number of localized states at specific locations. The Fano resonances are classified into two types: ones corresponding to single-quantum-dot states and ones corresponding to DQD states. It is noticeable that, for the DQD case, there is a splitting of the energies of the localized states resulting from the superposition of quantum states in the strain-induced DQD. Thus, like the singlet states of a DQD in a 2DEG~\cite{Hayashi2003,Petta2004,Gorman2005}, our DQD states are able to be treated as a two-level system in graphene, induced by elastic strain effects.

The existence of the singlet states indicates that there are two identical QDs in the nanobubble that are coupled to each other. To examine the possibility for a qubit application, we consider a device design with asymmetric p-n junctions as a means of tuning the electronic states of the DQDt. As shown in Fig. \ref{fig:model}, our system is composed of three domains which can be subjected to different electrostatic potentials. The potential profile is modeled as follows:
\begin{align}
U\left(x,y\right)=U_{0}\tanh{\left[\frac{x+0.5d\left(1+\tanh{y}\right)}{\xi}\right]},
\end{align}
where $d$ is the distance shift of the p-n junction via plunger gate electrode. For $d=0$, we have a straight line of the p-n junction interface, relating to the singlet states as displayed in Fig. \ref{fig:model}(c). On the other hand, as $d$ varies, the p-n junction interface becomes bent, and the couplings to the two QDs accordingly become asymmetric. As a result, the two QDs are no longer identical, and thus the singlet states also no longer exist [see Fig. \ref{fig:model}(d)].

\begin{figure}
\centering
\includegraphics[width = \linewidth]{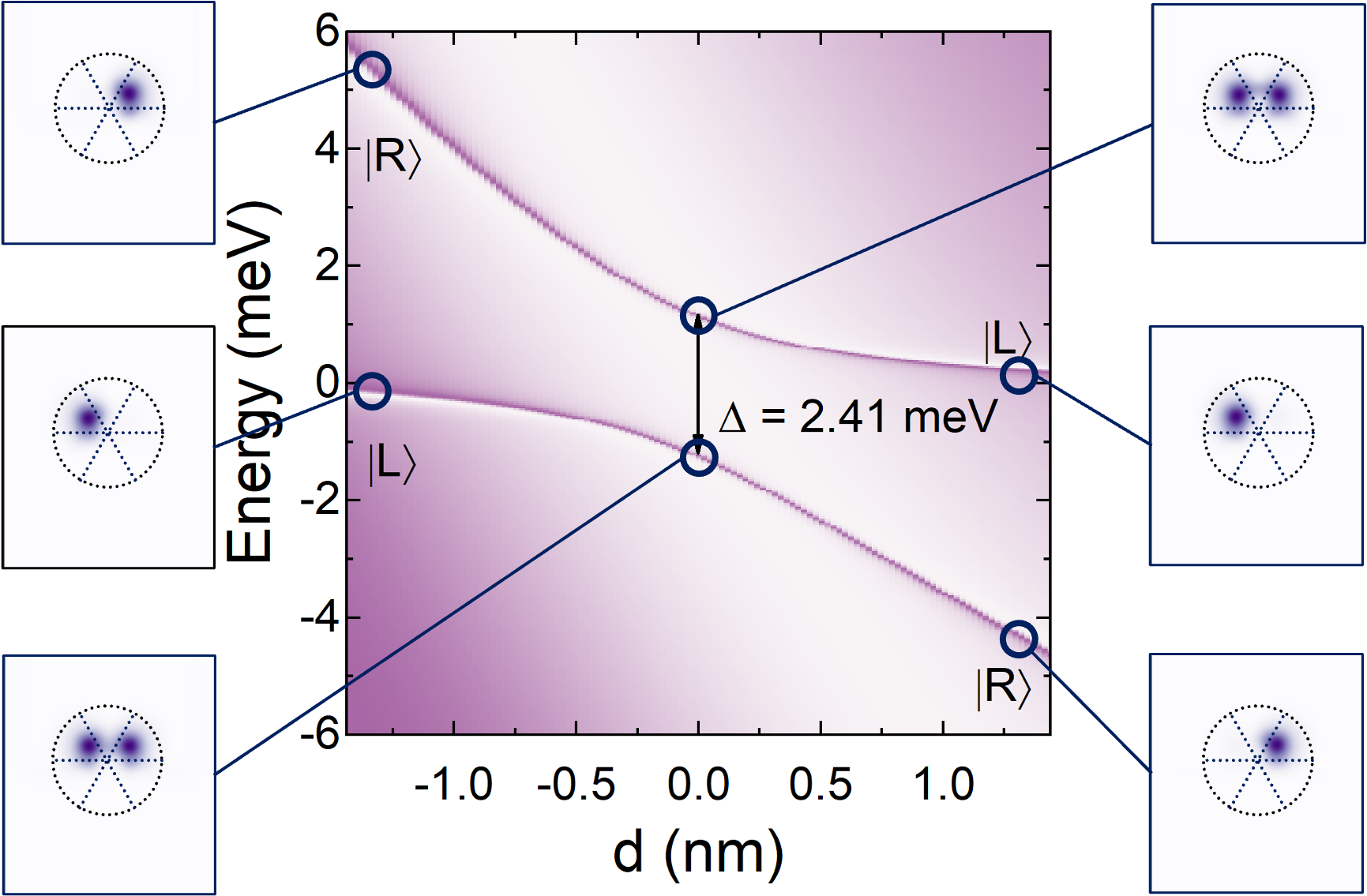}
\caption{Avoided crossing behavior of the graphene DQDt. The Fano resonance lines represent the localized states of the strain-induced QD in the nanobubble. Bright and dark colors indicate $2e^{2}/h$ and zero, respectively. Insets display the probability density distributions of interesting cases, denoted by open circles on the color map.}
\label{fig:avoided-crossing}
\end{figure}

Figure \ref{fig:avoided-crossing} shows conductance spectra in the vicinity of the singlet states for $x_{0}=27.25~a_{0}\simeq6.54$ nm with varying $d$. As expected, for $d=0$, we observe a clear avoided crossing behavior. At the avoided crossing point, the two-level system consists of symmetric and antisymmetric superpositions of two states,
\begin{align}
\left|\psi_{+}\right>=\alpha\left|L\right>+\beta\left|R\right>,\qquad \left|\psi_{-}\right>=\alpha\left|L\right>-\beta\left|R\right>,
\end{align}
where $\left|L\right>$ and $\left|R\right>$ indicate the ground states of QD1 and QD2, respectively. Let us notice that, at the avoided-crossing point, the $\left|\psi_{-}\right>$ state corresponds to the lower branch due to the $\pi$ Berry phase of graphene.

By decreasing $d$ from the avoided-crossing point, as displayed in Fig. \ref{fig:model}(d), QD2 becomes distant from the p-n junction interface, so that the electronic states of QD2 change with $d$. Accordingly, the two-level system continuously varies from the singlet states to two individual ground states. Since the coupling between the interface channel and QD1 remains unchanged by $d$, the lower branch of the two-level system converges on the $\left|L\right>$ state where only QD1 is allowed to be occupied by an electron, whereas the upper branch converges on the $\left|R\right>$ state where only QD2 is allowed to be occupied by an electron. On the other hand, by increasing $d$ from the avoided-crossing point, the upper and lower branches encounter $\left|L\right>$ and $\left|R\right>$ states, respectively.

Meanwhile, we calculate an energy splitting of $\Delta=2.41$ MeV at the avoided-crossing point, which is larger than other semiconductor QD qubits~\cite{Gorman2005,Goswami2007,Borselli2011,Nichol2017}. Such a large $\Delta$ value corresponds to $\Omega=\Delta/h=583.8$ GHz for the Landau-Zener transition at the avoided-crossing point. One can take advantage of the high-frequency Landau-Zener transition, allowing for sufficient numbers of qubit operations within finite coherence times on the order of hundreds of ns~\cite{Gorman2005}. In addition, as shown in Fig. \ref{fig:QDform}, there is another two-level system in the nanobubble with a larger energy splitting, $\Delta=4.2$ meV, allowing for higher frequency up to 1 THz.

\begin{figure}
\centering
\includegraphics[width = \linewidth]{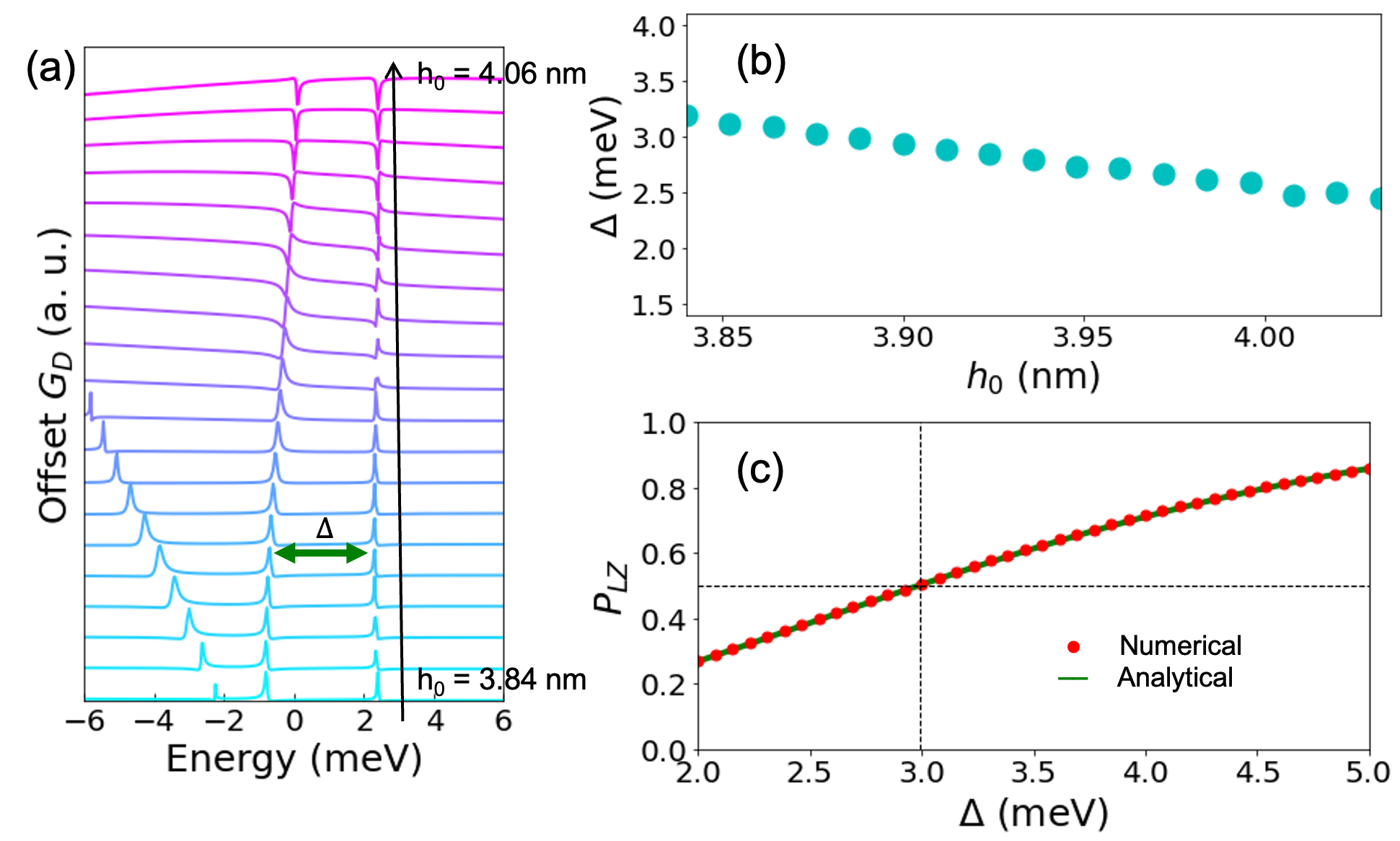}
\caption{(a) Conductance spectra near the avoided crossing point for various $h_{0}$ values. (b) Energy splitting $\Delta$ taken from the spacing between the Fano resonances shown in (a). (c) Landau--Zener transition probabilities at the avoided-crossing point as functions of $\Delta$ from both numerical and analytical calculations. Note that $\Delta$ is controllable via strain engineering of the nanobubble.}
\label{fig:tuning-gap}
\end{figure}

Next, we move our focus to the tunability of the Landau-Zener transition near the avoided-crossing point of the graphene DQD. As the two-level system approaches the avoided-crossing point, the coupling between the two QDs increases, such that the quantum state undergoes the Landau-Zener transition with the transition probability
\begin{align}
P_{LZ}=1-e^{-\frac{\pi\Delta^{2}}{2\hbar v}}, \label{eq:LZprob}
\end{align}
where $\Delta$ is the energy splitting at the avoided-crossing point and $v$ is the sweep speed of an energy evolution for qubit manipulation. Since the energy splitting $\Delta$ is associated with the coupling between QDs, it is obvious to expect that $\Delta$ is not constant but variable as the pseudo-magnetic field strength changes. In fact, unless the circular shape of the nanobubble is deformed, the characteristic profile of the pseudo-magnetic field is unchanged, while the strength of the pseudo-magnetic field varies by either $h_{0}$ or $\sigma$~\cite{Myoung2020}. Here, for simplicity, we focus on the effects of $h_{0}$ to see how $\Delta$ changes as the pseudo-magnetic field strength varies. Figure \ref{fig:tuning-gap} exhibits a clear dependence of $\Delta$ on $h_{0}$. The energy splitting $\Delta$ becomes smaller for stronger pseudo-magnetic fields, i.e., larger $h_{0}$ values. This behavior can be intuitively understood as follows: a stronger pseudo-magnetic field brings about a stronger confinement of Dirac fermions, which weakens the coupling between the QDs, which in turn results in smaller energy splitting at the avoided-crossing point.

Considering the strain-tunable energy splitting of the graphene DQD, it is straightforward to expected that the Landau-Zener transition can also be tuned by strain engineering. In particular, obtaining a half probability of the Landau-Zener transition is significant for quantum interference technology, like Landau--Zener--St\"{u}ckelberg interferometry. From Eq.~(\ref{eq:LZprob}), we find the required energy splitting for $P_{LZ}=0.5$,
\begin{align}
\Delta_{h}=\sqrt{\frac{2\ln{2}a\hbar\omega}{\pi}},
\end{align}
assuming that a time-dependent perturbation for qubit operation is a part of the sinusoidal time-evolution of the two-level system with amplitude $a$ and frequency $\omega$. Let us briefly note that $a\sim0.1$ V and $\omega\sim100$ GHz are practical values for graphene-based electronic device operation, so $a\omega$ is on the order of 10 eV$\cdot$GHz. For $\Delta_{h}=2.41$ , we need $a\omega=2$ eV$\cdot$GHz. Meanwhile, in order to see how the Landau-Zener transition probability is finely tunable, we perform a numerical calculation of the time-dependent two-level system for a different value, $a\omega=3.03$ eV$\cdot$GHz. The resulting Landau-Zener transition probability as a function of $\Delta$ is displayed in Fig. \ref{fig:tuning-gap}(c). We find $P_{LZ}=0.5$ when $\Delta=3$ eV, which corresponds to $h_{0}=3.89$ nm. If we consider that $\Delta$ is initially set to $2.41$ eV with $h_{0}=4.06$ nm, the 50\% Landau-Zener transition probability is acquired by finely tuning $h_{0}$ as much as 0.17 nm.

\begin{figure}
\centering
\includegraphics[width = 0.8\linewidth]{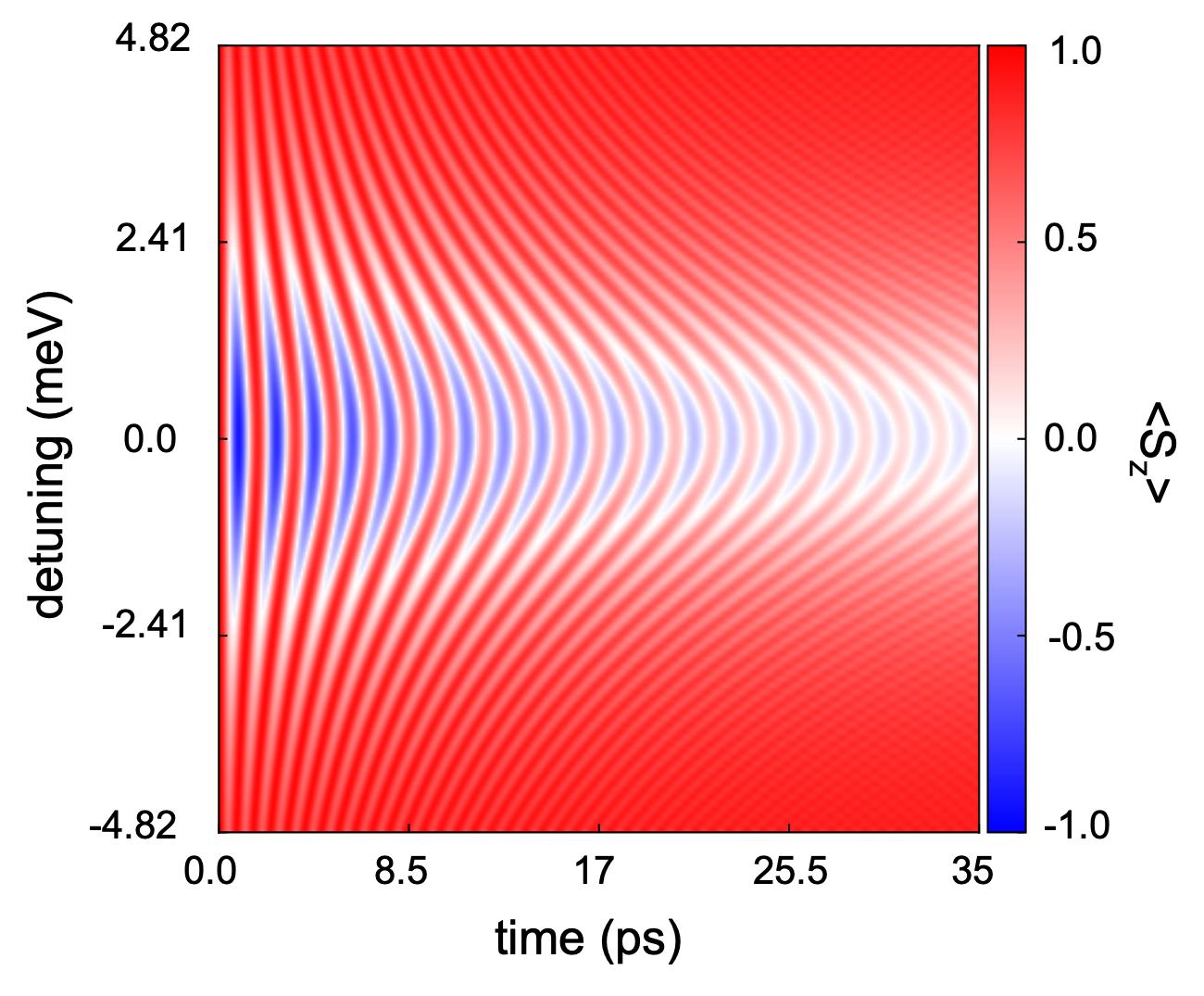}
\caption{Rabi oscillation map showing $S_{z}$ expectation values defined with the Bloch sphere concept. $\left<S_{z}\right>=1$ indicates that the two-level system resides in the $\left|\psi_{+}\right>$ state, i.e., the excited state. The zero detuning implies that two QDs in the nanobubbles are set to be identical for the two-level system formation at the avoided-crossing point.}
\label{fig:Rabi}
\end{figure}

We now examine the potential of our DQDs for graphene qubit applications by investigating the dynamics of the quantum states at the avoided-crossing point. Since the strain-induced DQDs in the nanobubbles are weakly but non-negligibly coupled with the interface channel at the p-n junction, we consider that the quantum states of the two-level system undergo dissipative interactions with the surroundings as a thermal bath. In order to describe such dissipative dynamics of the two-level system, we introduce the following effective Hamiltonian with the energy splitting $\Delta$ at the avoided-crossing point,
\begin{align}
H=\frac{\varepsilon}{2}\left(\sigma_{0}+\sigma_{z}\right)+\frac{\Delta}{2}\sigma_{x}
\end{align}
where $\varepsilon$ is the detuning energy. The dynamics of the quantum states with the dissipative interaction is governed by the Lindblad equation,
\begin{align}
\frac{d\rho}{dt}&=-\frac{i}{\hbar}\left[H,\rho\right]\nonumber\\
&+\left(\Gamma_{\varphi}\mathcal{D}_{\sigma_{z}}\left[\rho\right]+\Gamma\left(1-\overline{n}\right)\mathcal{D}_{\sigma_{-}}\left[\rho\right]+\Gamma\overline{n}\mathcal{D}_{\sigma_{+}}\left[\rho\right]\right),\label{eq:Lindblad}
\end{align}
where $\mathcal{D}_{L}\left[\rho\right]= L\rho L^{\dagger}- \frac{1}{2}\left\lbrace L^{\dagger}L, \rho\right\rbrace$ with $\left\lbrace A, B\right\rbrace$ being an anti-commutator between two operators $A,B$, $\overline{n}$ is the Fermi--Dirac distribution at temperature $T$ of the thermal bath, $\Gamma_{\varphi}$ is the dephasing rate of the quantum states, and $\Gamma$ is the relaxation rate as a result of the coupling between the DQD and the interface channel~\cite{BRE02}. Note that in absence of $\Gamma_{\varphi}$ and $\Gamma$, the two-level system becomes fully coherent. Due to the long phase relaxation time of graphene~\cite{TRAUZETTEL2007,Coish2004}, we assume that the dynamics of the quantum states is dominated by the dissipative coupling with the thermal bath. In this study, the relaxation rate due to the dissipation is extracted from the numerical results of the Fano resonances, $\Gamma\simeq 86$ GHz. Numerically solving Eq.~(\ref{eq:Lindblad}), we can obtain the expectation values of $\sigma_{z}$. Figure \ref{fig:Rabi} shows clear Rabi oscillation when zero detuning $\varepsilon=0$, undergoing attenuation as a consequence of the dissipative coupling. From Fig. \ref{fig:Rabi}, the period of the Rabi cycle is estimated to be 1.75 ps; this value is in a good agreement with $\Omega=\Delta/h=583.8$ GHz $\simeq \left(1.75~\mbox{ps}\right)^{-1}$.

Strain-induced graphene DQDs in nanobubbless are advantageous for two key reasons: a strain-tunability of the quantum states, and ultra-fast operation within the long coherence time of graphene. In conclusion, we expect that our results will help inaugurate strain-engineered qubits, further exploiting the properties of pristine graphene.

The authors would like to acknowledge the support of the National Research Foundation of Korea (NRF-2019R1F1A1051215), Project Code (IBS-R024-D1), and Chosun University (2020).

\bibliographystyle{apsrev4-1}
\bibliography{StrainQubit}

\end{document}